\documentclass{IEEEcsmag}

\usepackage[colorlinks,urlcolor=blue,linkcolor=blue,citecolor=blue]{hyperref}

\usepackage{tabularray, fontawesome}
\usepackage[normalem]{ulem}
\usepackage{rotating}
\usepackage{upmath}

\jvol{XX}
\jnum{XX}
\paper{8}
\jmonth{September}
\jname{IEEE Software}
\pubyear{2026}

\begin{document}

\sptitle{Department: Head}
\editor{Editor: Name, xxxx@email}

\title{AI Policies: Help or Hindrance? A Software Developer's Perspective}

\author{Samuel Ferino}
\affil{Monash University}

\author{Rashina Hoda}
\affil{Monash University}

\author{John Grundy}
\affil{Monash University}

\author{Christoph Treude}
\affil{Singapore Management University}

\author{Hashini Gunatilake}
\affil{Monash University}

\markboth{Department Head}{Paper title}

\begin{abstract}
AI policies introduced by software organisations to mitigate LLM-related risks such as sensitive information leaks and unauthorised usage are not useful if software developers do not engage with them. We draw on 19 software developer interviews to show how AI policies help and hinder developers. We suggest approaches to support managers and decision makers with a developer-centric approach to introducing AI policies.
\end{abstract}

\maketitle

\section{Introduction}

According to the 2025 Google Cloud ROI of AI report, 88\% of early adopters leveraging LLM-powered software report a positive return on investment, with 70\% citing immediate productivity gains across enterprise IT operations \cite{GoogleCloudROI:2025}.
However, security and risk concerns remain the primary barrier to scaling AI, as highlighted by the 2026 McKinsey AI Trust Maturity Survey of nearly 500 organisations \cite{McKinsey:2026}. Deloitte's 2026 State of AI Survey indicates that data privacy and security issues are the top concern among respondents ($73\%$), followed by legal, intellectual property, and regulatory compliance risks ($50\%$) \cite{Deloitte:2026}. Recent research findings underscore these concerns, demonstrating that LLMs remain vulnerable to security and privacy attacks such as jailbreaking and data poisoning \cite{das:2025}.

Despite these high operational and security stakes, organisational governance has fallen behind. The same Deloitte Survey reports that only 21\% of companies maintain a mature governance framework for autonomous AI agents, amplifying risks surrounding data privacy and centralised visibility, such as unmonitored employee usage of AI tools \cite{Deloitte:2026}.
Addressing these governance gaps requires organisations to introduce or update their AI usage policies, structures, and workflows. However, such structural adaptations can incur substantial costs in terms of time and financial resources and disrupt existing software development processes \cite{klotins:2022,giardino:2015}. Consequently, organisations must carefully balance risk mitigation with the impact of process changes on the humans working within those boundaries.

Neumann et al. \cite{neumann:2026} tried to address this gap by investigating how developers and agile coaches in three German organisations adopt GenAI tools, focusing on regulatory conditions. Their 17 semi-structured interviews reveal that a mismatch between top-down governance and bottom-up practices (e.g., prohibiting LLMs without providing practical alternatives) drives \textit{shadow IT} because policy is perceived as impractical. Similarly, Ronanki et al. \cite{ronanki:2026} investigated best practices for responsible adoption of LLMs through a multi-case study with three organisations involving developers, product owners, and scrum masters. Their findings cover seven trustworthy AI compliance recommendations, such as ``avoid task-misaligned fine-tuned LLMs'', ``restructure business processes to integrate LLMs'', and ``prioritise usefulness over strict accuracy''.  

Khojah et al. \cite{khojah:2025} investigated the implications of LLM policies in software organisations through the lens of engineering managers. Their interview study with 11 managers highlighted how organisations formulate these policies and the drivers behind them. However, while managerial perspectives shed light on policy design, understanding developers' perspectives is critical to understand the impact of LLM company policies in software development, since no amount of governance and policy will help if developers do not engage with them.

To capture a vital developer perspective, we interviewed 19 software developers across four regions in North America, Latin America, Europe, and Australasia. We sought to answer the question: 
     \textbf{How do organisational AI policies (on LLM-based tools) for software development impact software developers?}
Through our empirical findings, we provide insights into what helps and hinders software developers when it comes to AI policies. By illuminating the downstream impacts of policies, we intend to support IT managers, engineering leaders, and decision makers in navigating AI policies within their organisational context.

\section{Methodology}

We conducted a semi-structured interview study (approved by the Monash University Human Research Ethics Committee, Project ID: 44875) involving 19 software developers with experience of using LLMs for software development tasks. We advertised our study on our professional social media, LinkedIn and X (formerly Twitter). The recruitment focused on participants with experience using LLMs for software development tasks.

\begin{figure*}
  \centering
    \includegraphics[width=\linewidth]{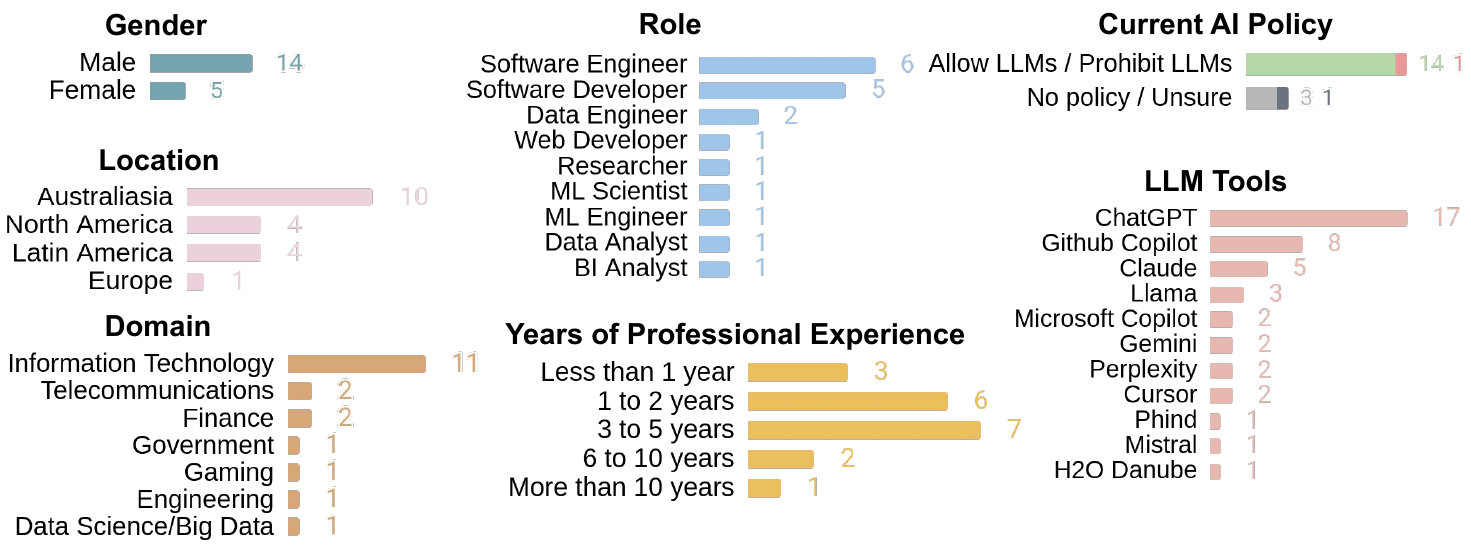}
  \caption{Demographics and Professional Background of Interview Participants.
  }
  \label{fig:demographics}
\end{figure*}

Participants completed a pre-interview questionnaire detailing their demographic and professional background (Figure \ref{fig:demographics}) and their company's AI policy regarding the use of LLMs for software development,  where P18 had a prohibitive policy at the time of the interview while P2, P12, and P16 reported having moved from prohibitive to permissive policies.

During the semi-structured interviews, ranging from 34 to 58 minutes, we sought details regarding their use of LLMs for software development including company policies by asking follow-up questions tailored to their context: 

\begin{itemize}
    \item \textit{``In the pre-interview questionnaire, you indicated that your organisation [has / does not have] an explicit policy regarding the use of LLMs for software development. Could you elaborate on this?} 
    \item \textit{When was it introduced, and by whom?}
    \item \textit{Do you know how it was developed or what prompted its introduction?}
    \item \textit{How has this policy—or lack thereof—affected you or your team? Has it been a help or a hindrance?''}. 

\end{itemize}

We analysed the interview transcripts using socio-technical grounded theory for data analysis (STGT4DA) \cite{hoda:2024}.  The first author conducted the data analysis, applying open coding, constant comparison, and memoing, while the co-authors reviewed the codes and memos throughout the process. The pre-interview questionnaire, the interview guide, and some de-identified examples of qualitative data analysis are available in our supplementary package\footnote{Supplementary file: \url{https://doi.org/10.5281/zenodo.22581906}} to support future researchers, in line with the ethical research approval governing this study.

\textbf{Limitations.} Nineteen practitioners is a modest sample size. Our sample size is limited regarding developer experiences of working under prohibitive AI policies. Our participant recruitment carries potential self-selection bias as a study limitation.

\section{Organisational Policies for Governing LLM Adoption}

As shown in Figures \ref{fig:demographics} and \ref{fig:findings}, interview participants' descriptions revealed four distinct AI organisational policies: allow LLMs, prohibit LLMs, no policy, and unsure. Organisations permitting LLMs either allow unrestricted access or limit usage to specific licensed models, whereas prohibiting organisations ban LLM use entirely in the workplace. Cases with no policy lack explicit guidelines on LLM usage, while uncertainty reflects instance where an organisation's stance on LLM adoption remains unknown to employee.

Our interview participants described their understanding of the \textbf{motivations} underlying organisational decisions to prohibit, permit, or defer LLM use (Figure \ref{fig:findings}). 

\textbf{Prohibiting LLM use.} Participants (P2 and P18) explained how their organisations--motivated by the desire to prevent potential legal and security issues--exhibited a risk averse approach and prohibited their developers from using LLMs. Customers may lose trust on how the organisation manages their data is another motivation (P18). 

\begin{quote}
\small
\textit{``Their policy was just to avoid using ChatGPT or any other LLMs. Their concern was more to do with the ownership of the code that was being generated, and how it might violate trademarks and violate the licenses. So they really didn't want any legal trouble."} - P2 (Software Engineer, IT). 
\end{quote}

Despite this, shadow AI usage for code understanding is still employed by P18 - e.g., \textit{``I'm not really using ChatGPT for my office coding work directly, but for [code] understanding''}.

\textbf{Enabling controlled LLM adoption.} Organisations may treat LLMs as just another tool, and developers are clearly informed about their responsibilities regarding accountability (P8). Developers can use AI assistance during coding-related activities, but they may also need to explicitly declare it (P17).

Several participants (P2, P7-P8, P14-P16) shared how organisations navigate concerns related to security, code ownership, and company reputation when developing permissive policies for LLM use. 

To mitigate security issues, P14 reported that their organisation provides \textit{licensed LLMs}, which may allow the organisation to monitor employees' usage of LLMs. P7 and P16 mention that their organisation defines \textit{prompting guidelines} and relies on developer prompting monitoring: 

\begin{quote}
\small
\textit{``I did a training on using AI in the workplace, and they said they were building a company policy, because at the moment they're just relying on their workers to not put company data into the prompts"} - P7 (BI Analyst, Telecommunications). 
\end{quote}

Additionally, the organisation may allow developers to use LLMs but \textit{restrict full integration} with internal systems.
\begin{quote}
\small
\textit{``In a Power BI Pro account, there's an integrated Copilot that you can use, and it's very handy [...] you can use it to create insights for visuals [..] I am assuming it's because of security reasons [that] we've [in my company decided] turned this function off. I'm not sure if there would be a method to use it securely"} - P7. 
\end{quote}

P15 explained that they communicate with LLMs via private APIs while P2 and P8 argued that developers could be permitted to use LLMs after their organisations resolve code ownership concerns.

\begin{quote}
\small
\textit{``As long as they sort of fix the whole ownership, like the licensing issues that are with LLMs, I think it's okay using LLMs, because it's just another tool that will help you do your job"} - P2 (Software Engineer, IT).
\end{quote}

However, these security mitigation strategies might not work for every situation. For this reason, some organisations go the extra mile to allow LLMs - e.g., \textit{``But other teams, like the data engineering team, would like to have additional security or safety procedures on top of that. So they were working on some kind of additional policies and guidelines to further safeguard the whole LLM ecosystem."} - P15 (Researcher, Engineering). P7 and P15 mention that their organisation decided to put effort into clearly defining the allowed boundaries - e.g.,  \textit{``they were working on the [AI] policy to define it better and really clarify what specific data should not be passed into which LLMs or which tools to use."} - P15 (Researcher, Engineering).

\textbf{Deferring policy development.}  Although LLM policies can support organisational processes and help to mitigate LLM risks, the focus on product release for early-stage organisations may surpass security concerns, and introducing policies may slow down their developers (P5). This may also apply to organisations where the product does not differ that much from competitors (P13). Organisations may still be trying to understand the business value of LLMs, and drafting an AI policy is not a priority for them - e.g., \textit{``AI is quite new, and I think our company has yet to write a policy on it. Maybe they will [write an AI policy] at some point in time. But in general, we're exploring ways that we can use AI in our platform, like we want to integrate some kind of AI into our platform"} - P13 (Software Developer, IT).

\textbf{Incremental and evolving approach.} We found an incremental and evolving approach to AI policy development.  Some organisations are gradually shifting from complete prohibition to permitting the use of licensed  (P12, P16).

\begin{quote}
\small
\textit{``The first policy they put in place was to communicate company-wide that you cannot use ChatGPT... and we had to wait for the licenses to come... I think they've been doing the licensing quite quickly, because we got access... so that there weren't any issues of data leak, because it's a huge concern for the company."} - P16 (Software Engineer, Gaming).    
\end{quote}

The product domain may also influence how organisations approach LLMs, e.g., \textit{``It's always going to be depending on the sensitivity of what you're doing. If a bank ends up doing a full app with an LLM, I won't trust that bank anymore."} - P16 (Software Engineer, Gaming).

\section{Help or Hindrance? The Developer Impact}

\begin{figure*}
  \centering
    \includegraphics[width=\linewidth]{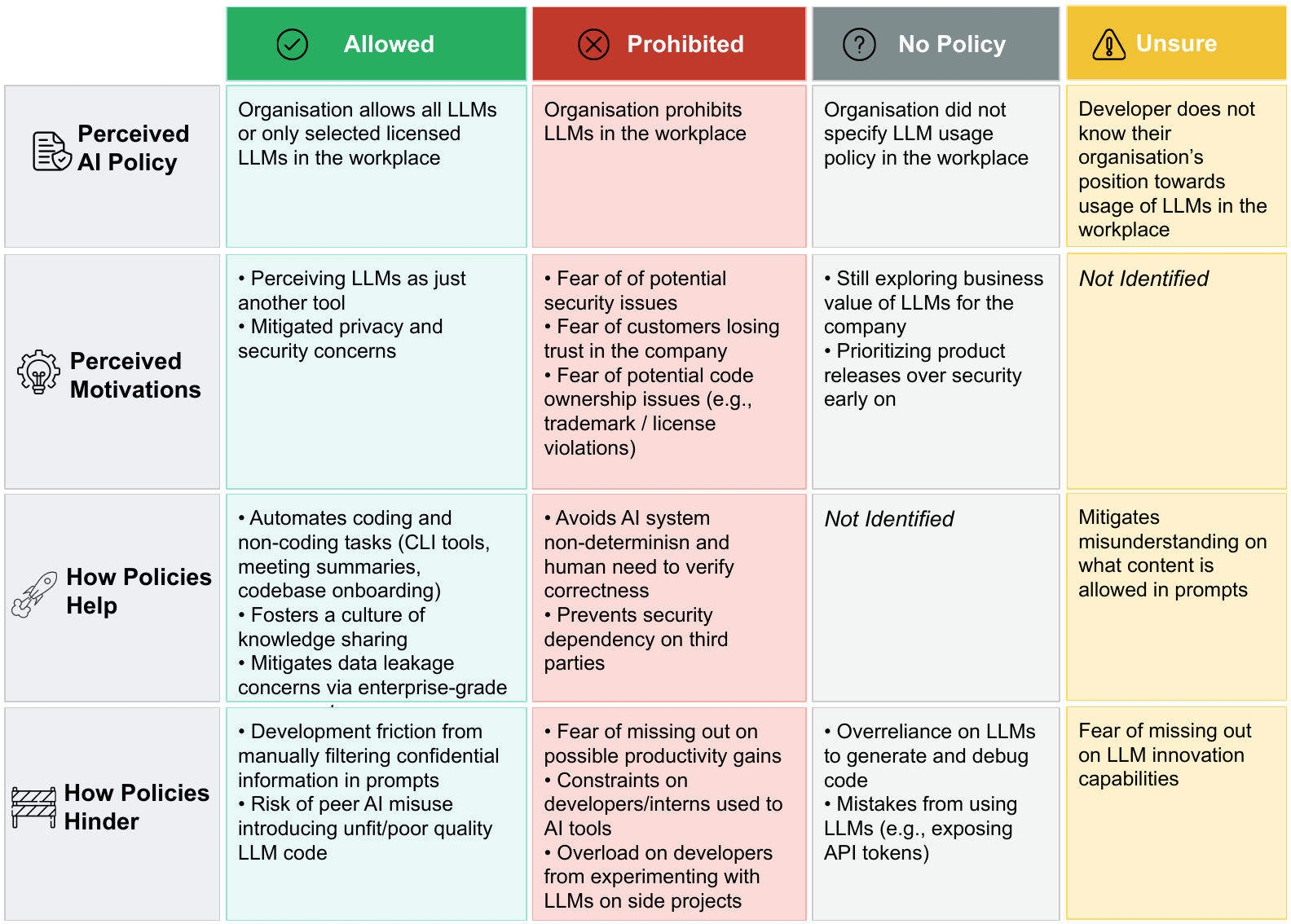}
  \caption{Summary of AI policies, perceived motivations, and positive \& negative impact on developers.}
  \label{fig:findings}
\end{figure*}

The two-sided nature of permissive and prohibitive policies can be a help or hindrance to software developers, and developers shared their perceptions about the absence of clear policies. Figure \ref{fig:findings} summarises the AI policies, positive and negative impact on developers.

\subsection{How do Policies Help?}  

While some benefits of \textit{permissive} policies, such as enabling workflows, may seem intuitive, our analysis reveals that \textit{prohibitive} policies were also seen to be advantageous to developers and organisations in some ways.

\textbf{Helpful When Allowed.} The benefit of \textit{permissive} policies is having access to enterprise grade LLMs. Going beyond mere permission, P14 and P19 argue that providing licensed LLMs actively mitigates concerns regarding data leakage and unauthorised usage of data for LLM training. 
\begin{quote}
\small
\textit{``The [LLM] tools we have used, those are actually for enterprise use. They have [an] enterprise [level service] agreement between my company and them, so those [data] are not actually used for training"} P19 (Machine Learning Scientist, IT).
\end{quote}

Having access to LLMs it also enables automating coding related as well as non-coding related tasks (P10, P11, P14, and P15). For tasks beyond coding, LLMs were seen to help with meeting summarisation (e.g., with Microsoft Copilot) and a smoother company onboarding process by supporting developers in familiarising themselves with the codebase. 

\begin{quote}
\small
\textit{``A lot of times, you're gonna miss some of the meetings. There'll be a lot of times where some meetings could have been an email. So in that case, if I'm late to a meeting, I will ask Copilot to generate the summary of what I've missed"} -- (P14 Data Engineer, Telecommunications) 
\end{quote}

Access to LLMs was also seen to help foster a culture of knowledge sharing (P4, P12, and P17).
\begin{quote}
\small
\textit{``My senior engineer said to me: `hey, you know what I'm using [is] Copilot'. [...] and then I was: `What's the difference [between ChatGPT and GitHub Copilot]?' And he was like: `Oh, yeah, it [GitHub Copilot] reads the whole codebase. And it knows what you ask for. [...] let's try it out', and he just did a prompt there and  [the response was] good."} - P12 (Software Engineer, Finance)
\end{quote}

\textbf{Helpful When Prohibited.} While the benefits associated with permissive AI policies were somewhat expected, one developer (P18) said that prohibitive AI policies helped her organisation to avoid non-determinism in their systems and the constant concern that the system response may not always be correct. It was also seen to prevent security exposure and dependency on third parties. 
\begin{quote}
\small
\textit{``When it comes to AI, you lose security and confidentiality"} - P18 (Software Developer, Finance).
\end{quote}

\textbf{Helpful When Unsure.} When asked about how his manager could help him, P1 expressed his desire for clear boundaries - e.g., \textit{``you have to indicate the limits, indicate what can be done and what can't be done''} - P1.

Future research could validate these observations across a larger population of IT professionals working on projects under LLM-prohibitive policies (e.g., highly confidential projects, high risk domains).

\subsection{How do Policies Hinder?} 

\textit{Prohibitive} policies were seen to cause hindrances to developer experience, e.g., on account of missed opportunities. Perhaps somewhat surprisingly, \textit{permissive} policies were also associated with being a hindrance to developers.

\textbf{Hindrance When Prohibited.} A number of challenges were discussed in the context of \textit{prohibitive} policies (P2, P11-P12, P15-P16, P18). Without LLM assistance in coding related tasks, P15 and P18 anticipated a downturn in productivity and innovation - e.g., \textit{``It would take much longer to do something you [that] can do with ChatGPT or Claude nowadays [...]  mostly debugging, sometimes even writing code"} - P15 (Researcher, Engineering). 

P16 observed that coworkers who moved from organisations that allow LLM usage to others that prohibit them struggled with coding-related tasks:

\begin{quote}
\small
\textit{``We have a lot of interns coming over [...] we have a couple of masters who [are used to using] GPT or Copilot a lot to code, and [now] they cannot inside the company. So, it is a constraint on them [not being able to use LLMs]."} - P16 (Software Engineer, Gaming)    
\end{quote}

P2 and P12 reported starting to explore LLM usage on the side, through personal projects, due to organisational restrictions: \textit{``to be honest, I started using this outside of work rather than at work, because we were told to not use it because of all the licensing issues..."} - P2 (Software Developer, IT). These side projects can overload developers, resulting in accumulating stress and damage to their wellbeing \cite{silva:2025}.

\textbf{Hindrance When Allowed.}  Under permissive policies, developers may encounter unintended development friction and expose teams to the risks of peer AI misuse (P7, P10, P15-P17). While navigating security issues, P7, P15, and P17 need to go through the hassle of filtering confidential information when prompting.  P16 observed that security concerns are more common among experienced developers, who better understand the implications of code leaks, and might prefer to avoid LLMs even when allowed. 

P10 shows concerns about developers using LLMs to introduce unfit quality LLM-generated code into the codebase - e.g., 

\begin{quote}
\small
\textit{``[LLMs] might not always choose the best software engineering practices [...] if something is really popular, [it] doesn't mean it's the best [for your project]. [It means] it's just popular. A lot of code on GitHub is done by people who are just trying to learn things, and they might not actually have the experience with that, or know if that [code] scales well"} - P10 (Software Engineer, IT).
\end{quote}

\textbf{Hindrance when No Policy.} Drawing on participants' experiences working without clear organisational policies, we found the following perceived challenges: irresponsible LLM usage, such as \textsl{vibe coding} in a way that developers overrely on LLMs to generate and debug code, and potential mistakes that expose sensitive information (e.g., API tokens) in prompts (P5, P13).

\textbf{Hindrance when Unsure.} Our participant P1 reports challenges he faced while not aware of his organisation's policy. For instance, P1 feels uncomfortable with the situation of putting proprietary code into an LLM without fully understanding the level of sensitivity.

\begin{quote}
\small
    \textit{``A lot of times when we work for a private company, we feel very uncomfortable sending code; you can't put private code into ChatGPT or that kind of thing. I'd say it's a grey area of understanding what you can and can't send to them"} - P1 (Software Developer, IT).
\end{quote}

There are also concerns about losing LLM innovative potential due to prohibitive policies - e.g., \textit{``We'd lose so much in terms of time or innovation. It's already there; what are we going to do?''} - P1.

\section{Recommendations for Evolving Policies to Better Support Developers}

\subsection{Implications for Practice}

We organise our discussion around strategies to mitigate policy hindrances and an LLM policy progression checklist.

\textbf{Mitigating Policy Hindrances.} We suggest the following approaches grounded in the literature to manage the hindrances:

\begin{itemize}
    \item Seamless security layers through automated prompt sanitisation and traditional code quality check workflows can mitigate development friction arising from developers having to manually filter confidential information in prompts, the risks of peer AI misuse introducing poor-quality LLM code, and the risks of sensitive information being leaked to the LLMs \cite{owaspllm02:2025, owaspllm07:2025}. 

    \item Providing a sandbox (e.g., CodeRabbit\footnote{\url{https://docs.coderabbit.ai/code/environments}}) as a safe and isolated development environment where developers who are accustomed to using AI tools for software development can work with fewer constraints \cite{pombo:2026}.

    \item Regular meetings focused on sharing experiences using AI tools (e.g., new features) can provide a welcoming environment for developers who are experimenting with LLMs on side projects, as well as for those who are afraid of missing out on possible productivity gains from using LLMs \cite{altork:2026}.
    
\end{itemize}

\textbf{LLM Policy Progression Checklist.} To assist managerial decision-making when transitioning from prohibitive to permissive policies, we suggest using the checklist below based on our empirical findings. Answering \textit{yes} to most of these questions provides a strong indication that permissive policies could be adopted; otherwise, we recommend considering delaying deployment of new policies while addressing any gaps.

\begin{itemize}
    \item Are there built-in safeguards and guardrails in the LLMs you are considering for your developers to use?
    \item Have the developers gone through responsible AI usage training?
    \item Is there a clear definition of what constitutes sensitive information across the organisation, including customer information?
    \item Are developers fully informed of the risks of using LLMs for development?
    \item Are developers aware of their selected LLM services' privacy policy?
    \item Are the code ownership responsibilities clear?
    \item Is AI usage in code properly declared?
    \item Has the development team fully transitioned to LLM data protection services?
    \item Has the development team been introduced to any prompting guidelines?
\end{itemize}

\subsection{Recommendations for Future Research}

We suggest the following future investigations in human-centric AI governance:

\begin{itemize}
  
    \item How accurately do software developers interpret organisational AI policies? What are the primary knowledge gaps between manager intent and developer understanding?

    \item To what extent can organisations monitor responsible AI adherence during software workflows without compromising developer privacy?

    \item What \textit{shadow IT} practices do developers adopt under restrictive or ambiguous AI policies? How can policy design mitigate them?

    \item To what extent do prohibitive AI policies drive developers to experiment with LLMs on personal side projects? How does this contribute to developer stress and burnout?

    \item How do policies  impact talent retention and developer satisfaction? Does prohibitive policies correlate with increases in organisational turnover?

    \item What are the specific workarounds developers adopt to navigate and mitigate LLM-related security risks, enabling permissive policies?
\end{itemize}

\section{Conclusion}

Organisations are facing the challenge of deciding how to leverage LLM tools in software development while mitigating organisational risks. In this study, we addressed this through the lens of software developers by examining their experiences and perceptions of how organisational AI policies impact them. We identified a governance progression across some participants' organisations, maturing from strict LLM prohibitions toward permissive access via licensed LLMs. We also found a two-sided nature of permissive and prohibitive policies, both serving as a help and a hindrance to software developers. To assist IT managers, we propose actionable recommendations to mitigate the hindrances and a LLM policy progression checklist. The main takeaway is that, aside from exceptional cases, AI governance could be treated as an evolving lifecycle, transitioning from prohibitive to permissive policies as soon as gaps are addressed.

\section{Acknowledgments}
Ferino is supported by a Faculty of IT Postgraduate Scholarship. We express profound gratitude to all participants who took part in this study. We thank Humphrey Obie for feedback on this research and Nimmi Weeraddana and Haoyu Gao for helping with participant recruitment.

\bibliographystyle{IEEEtran}
\bibliography{bibliography}

\begin{IEEEbiography}{Samuel Ferino}{\!} is a PhD candidate at Faculty of Information Technology, Monash University, Australia. His research interests include LLMs for software development, and education in software engineering. Ferino received his Master in System and Computing  from the Federal University of Rio Grande do Norte, Brazil. Contact him at samuel.demouraferino@monash.edu.
\end{IEEEbiography}

\begin{IEEEbiography}{Rashina Hoda}{\!} is a Professor of Software Engineering at Monash University, Australia. Her research focuses on the human and socio-technical aspects of SE at the intersections of AI and digital health. She was named the 2025 Top Researcher in Software Systems in Australia by The Australian. In her 2024 Springer book ``Qualitative Research with Socio-Technical Grounded Theory'', she presents a modern socio-technical variation to the Grounded Theory methods for SE. Rashina serves on the ICSE steering committee and as a guest editor for TOSEM's special issue on human-AI collaboration in SE. More about her on www.rashina.com Contact her at rashina.hoda@monash.edu.
\end{IEEEbiography}

\begin{IEEEbiography}{John Grundy,}{\!} is an Australian Laureate fellow and a professor of software engineering at Monash University, Melbourne, Australia. His current interests include domain–specific visual languages, model–driven engineering, large-scale systems engineering, and software engineering education. Grundy received his PhD in computer science from University of Auckland, New Zealand. He is a Fellow of the IEEE, Engineers Australia and ASE. He
is an associate editor of the IEEE Transactions
on Software Engineering, the Automated Software
Engineering Journal, and IEEE Software. More details about his research can be found at https://sites.google.com/site/johncgrundy/. Contact him at john.grundy@monash.edu.
\end{IEEEbiography}

\begin{IEEEbiography}{Christoph Treude,}{\!} is an Associate Professor of Computer Science at Singapore Management University. He has authored over 150 scientific articles with more than 250 coauthors. His work has received recognition, including an ARC Discovery Early Career Research Award (2018-2020) and funding from industry leaders such as Google, Facebook, and DST. Treude
has received four best paper awards, including two ACM SIGSOFT Distinguished Paper Awards. Currently, Treude serves on the Editorial Boards of the IEEE Transactions
on Software Engineering, the Springer Journal on Empirical Software Engineering, and the Wiley Journal of Software: Evolution and Process. He also holds the role of Open Science Editor for the Elsevier Journal of Systems and Software. He has chaired conferences such as ICSME 2020, ICPC 2023, and TechDebt 2023 and regularly participates in software engineering conference program committees.
\end{IEEEbiography}

\begin{IEEEbiography}{Hashini Gunatilake,}{\!} is a Research Fellow at Monash University,
Melbourne, Australia. She received her PhD in information technology from Monash University, Australia. Prior to her PhD, she was in the software industry. Her research interests are software engineering, human, social \& technical aspects, human-AI interaction, agile methodology, data visualisation. More details of her research can be found at https://hashinig.com. Contact her at hashini.gunatilake@monash.edu.
\end{IEEEbiography}

\end{document}